\documentclass[preprint]{revtex4}

\usepackage{graphicx}%
\usepackage{dcolumn}
\usepackage{amsmath}

\makeatletter
\def\btt#1{\texttt{\@backslashchar#1}}%
\DeclareRobustCommand\bblash{\btt{\@backslashchar}}%
\makeatother

\begin{document}

\title{Kantowski-Sachs universe cannot be closed}

\author{Xin-zhou Li}\email{kychz@shtu.edu.cn}

\author{Jian-gang Hao}

\affiliation{ Shanghai United Center for Astrophysics, Shanghai
Normal University, 100 Guilin Road, Shanghai 200234 , China}%

\date{\today}

\begin{abstract}
Abstract : In this paper, by analyzing the instability against
spatially homogeneous and anisotropic perturbations of the
Kantowski-Sachs-type during different cosmological epoch, we show
that it is a theoretical consequence of the general relativity
that the KS universe must be open or flat if it underwent the
matter dominated and/or radiation dominated era in its past
evolution, which theoretically confirms the flatness of our
observable universe.
\end{abstract}
\pacs{98.80.-k, 95.30.Sf}

\maketitle

\vspace{0.4cm}

The observations on the cosmic microwave background(CMB)
anisotropy and on the spatial distribution of galaxies on large
scale are believed to have a cosmological origin and it would be
very difficult to explain their existence and their isotropy if
the hypothesis of the spatial homogeneity and isotropy of the
universe were not valid to a very good approximation on large
scale. There is a connection between isotropy and spatial
homogeneity: unless a fundamental observer occupies a special
position in the universe, the isotropy implies spatial
homogeneity. The assumption of spatial homogeneity and isotropy of
space determines the metric completely except the sign of the
curvature. Considerably more freedom is left if one assumes only
spatial homogeneity of space\cite{one,two}. All homogeneous
cosmologies fall into two classes: the Bianchi models, for which
the isometry group admits a 3-dimensional simply transitive
subgroup and the Kantowski-Sachs models, for which the isometry
subgroup is neither simply transitive nor admits a simply
transitive subgroup\cite{three}. There exist nine Bianchi types
and, correspondingly, nine Bianchi cosmologies and each class has
subclasses with extra symmetries. The Bianchi types are in general
anisotropic so that they do not have all spatial directions at a
point being equivalent. Many authors have addressed the evolution
of spatially homogeneous cosmological models\cite{four}. Spatially
homogeneous cosmological models with a positive cosmological
constant are investigated\cite{five}. Exact string cosmological
solutions have been found for the Kantowski-Sachs model by Barrow
and Dabrowski\cite{six}. Reula\cite{seven} showed that all small
enough non-linear perturbations decay exponentially during
expanding phases of flat homogeneous cosmologies. Recently, Barrow
et al.\cite{eight} showed that the Einstein static model is
unstable to spatially inhomogeneous gravitational wave
perturbations within the Bianchi type IX class of spatially
homogeneous universes.

Recently, the Wilkinson Microwave Anisotropy Probe(WMAP) has
provided high resolution CMB data\cite{nine,ten,eleven} for
cosmology. Among the interesting conclusions that have been
reached from this data are constraints on present value $\Omega_0$
of the total density parameter of the universe. The new results
indicate that while the universe is close to being flat
$\Omega_0\simeq 1$, a closed universe is marginally preferred:
$\Omega_0>1$\cite{nine}. Especially, with a prior on the Hubble
constant, one gets that $\Omega_0=1.03\pm 0.05$ at 95 percent
confidence level, while combining WMAP data with SNeIa leads to
$\Omega_0=1.02\pm 0.04$ or to $\Omega_0=1.02\pm 0.02$ respectively
without and with a prior on the Hubble parameter. The latter may
be regarded as the present best estimate of this parameter. In
this paper, we study the instability against spatially homogeneous
and anisotropic perturbations, which can be used as probe of
curvature of space. We show that the anisotropy will not increase
when the expansion rate is greater than certain values while it
will increase when the expansion rate is less than that value or
the universe is contracting. We find that the radiation dominated
and matter dominated era, which correspond to the scale factor
$a(t)\sim t^{1/2}$ and $a(t)\sim t^{2/3}$, will not produce
significant anisotropy when and only when the universe is
spatially flat or open. In other words, we find a connection
between isotropy and non-closeness: \emph{the closed universe is
unstable against spatially homogeneous and anisotropic
perturbations of the Kantowski-Sachs-type during the radiation
dominated and matter dominated era}.

In this paper, the cosmological anisotropy is the whole anisotropy
of the scale factor and is different from CMB anisotropy. Why the
cosmological anisotropy is still very tiny today begs for
explanation. Therefore, it is very interesting whether degree of
cosmological anisotropy will increase during the evolution of
universe. We begin with the line element of anisotropic spacetime
\begin{equation}\label{metric}
ds^2=-dt^2+a(t)^2[(1+\delta)^2d\chi^2+d\theta^2+S^2(\theta)d\phi^2]
\end{equation}

\noindent where

\begin{equation}\label{s}
 S(\theta)=
  \begin{cases}
    \sin\theta & for \hspace{1cm} k=1, \\
    \theta & for \hspace{1cm} k=0,\\
    \sinh\theta & for \hspace{1cm} k=-1
  \end{cases}
\end{equation}

\noindent and $k=0$ and -1 are just axisymmetric Bianchi type I
and III universe while $k=1$ model, or the closed anisotropic
universe model, is referred to as the Kantowski-Sachs universe.
Although only the closed models fall outside of the Bianchi
classification, one can generally refer to them all as
Kantowski-Sachs-like models for convenience\cite{six}. In this
paper, we analyze instability against spatially homogeneous and
anisotropic perturbations within the Kantowski-Sachs-like models.
The Einstein equations correspond to the above setup are:
\begin{equation}\label{einst1}
 3H^2+\frac{2H\dot{\delta}}{1+\delta}+\frac{k}{a^2}=\kappa\rho
\end{equation}

\begin{equation}\label{einst2}
 2\frac{\ddot{a}}{a}+H^2+\frac{k}{a^2}=-\kappa p
\end{equation}

\begin{equation}\label{einst3}
 2\frac{\ddot{a}}{a}+H^2+\frac{3H\dot{\delta}}{1+\delta}+
 \frac{\ddot{\delta}}{1+\delta}=-\kappa p
\end{equation}

\noindent where $\kappa=8\pi G$, $p$ and $\rho$ are the pressure
and energy density of the perfect fluid respectively. The energy
conservation of the perfect fluid is expressed as
\begin{equation}\label{conservation}
\frac{d \rho}{dt}=-(3H+\frac{\dot{\delta}}{1+\delta})(\rho+p)
\end{equation}

\noindent Substitute Eq.(\ref{einst3}) into Eq.(\ref{einst2}), we
have

\begin{equation}\label{deltaeq}
\ddot{\delta}+3\frac{\dot{a}}{a}\dot{\delta}-\frac{k}{a^2}(1+\delta)=0
\end{equation}

It does not lose generality that we assume $\delta(t_0)\geq 0$ for
definiteness. In the following, we investigate the generic
evolution of the anisotropy in the power law expansion and
exponential expansion. Fortunately, Eq.(\ref{deltaeq}) can be
solved exactly in above cases. For the power law expansion
$a(t)=a(\frac{t}{t_0})^q$, the solution for Eq.(\ref{deltaeq}) are

\noindent Case(i): $q=0$
\begin{equation}\label{solution1}
\delta(t)=
  \begin{cases}
    \frac{1+\delta_0+a_0\dot{\delta_0}}{2}\exp(\frac{t-t_0}{a_0})+
     \frac{1+\delta_0-a_0\dot{\delta_0}}{2}\exp(\frac{t_0-t}{a_0})-1& for \hspace{1cm} k=+1, \\
    \dot{\delta_0}t+\delta_0-t_0\dot{\delta_0} & for \hspace{1cm} k=0, \\
    [(1+\delta_0)\cos(\frac{t_0}{a_0})-a_0\dot{\delta_0}\sin(\frac{t_0}{a_0})]\cos(\frac{t}{a_0})+
     [(1+\delta_0)\sin(\frac{t_0}{a_0})+a_0\dot{\delta_0}\cos(\frac{t_0}{a_0})]\sin(\frac{t}{a_0})-1
     & for \hspace{1cm} k=-1
  \end{cases}
\end{equation}

\noindent where $\delta_0\equiv\delta(t_0)$ and
$\dot{\delta_0}\equiv\dot{\delta}(t_0)$.

\noindent Case(ii): $k=0$ and $q\neq 0$
\begin{equation}\label{solution2}
\delta(t)=
  \begin{cases}
    \frac{t_0^{3q}\dot{\delta_0}}{t^{3q-1}}+\delta_0-\frac{\dot{\delta_0}t_0}{1-3q}
     & for \hspace{1cm} q\neq 1/3, \\
   t_0\dot{\delta_0}\ln(\frac{t}{t_0})+\delta_0 & for \hspace{1cm} q=1/3
  \end{cases}
\end{equation}

\noindent Case(iii): $k\neq 0$ and $q=1$
\begin{equation}\label{solution3}
 \delta(t)=C_1t^{-1+\sqrt{1+\frac{kt_0^2}{a_0^2}}}+C_2t^{-1-\sqrt{1+\frac{kt_0^2}{a_0^2}}}-1
\end{equation}

\noindent where integral constant $C_1$ and $C_2$ can be fixed by
initial values $\delta_0$ and $\dot{\delta_0}$.

\noindent Case(iv): $k\neq 0$ and $q\neq 0, 1$
\begin{equation}\label{solution4}
 \delta(t)=(\frac{t}{t_0})^{\frac{1-3q}{2}}\left[C_1Z_{\nu}^{(1)}\left(\frac{\sqrt{-k}t_0}{a_0(1-q)}
 (\frac{t}{t_0})^{1-q}\right)+C_2Z_{\nu}^{(2)}\left(\frac{\sqrt{-k}t_0}{a_0(1-q)}
 (\frac{t}{t_0})^{1-q}\right)\right]-1
\end{equation}

\noindent where $\nu=\frac{1-3q}{2(1-q)}$, $C_1$ and $C_2$ are
integral constants, $Z_{\nu}^{(1)}$ and $Z_{\nu}^{(2)}$ are Bessel
functions for $k=-1$ and the modified Bessel functions for $k=+1$
respectively, which can be fixed by the initial values. For
example, if $k=-1$ and $\nu\neq integer$, we have

\begin{equation}\label{c1forp}
C_1=\frac{t_0^{\frac{3\nu}{3-2\nu}}\left[(3-2\nu)^2(1+\delta_0)Z_{1-\nu}(x_0)
+4a_0\dot{\delta_0}Z_{-\nu}(x_0)\right]}{(3-2\nu)^2[Z_{\nu-1}(x_0)Z_{-\nu}(x_0)+Z_{1-\nu}(x_0)Z_{\nu}(x_0)]}
\end{equation}

\begin{equation}\label{c2forp}
C_2=\frac{t_0^{\frac{3\nu}{3-2\nu}}\left[(3-2\nu)^2(1+\delta_0)Z_{\nu-1}(x_0)
+4a_0\dot{\delta_0}Z_{\nu}(x_0)\right]}{(3-2\nu)^2[Z_{\nu-1}(x_0)Z_{-\nu}(x_0)+Z_{1-\nu}(x_0)Z_{\nu}(x_0)]}
\end{equation}

\noindent where $x_0=\frac{\sqrt{-k}t_0}{(1-q)a_0}$

For the exponentially expansion $a(t)=a_0\exp[H(t-t_0)]$, the
solutions of Eq.(\ref{deltaeq}) are as follows

\begin{equation}\label{expsolution}
\delta(t)=C_1\exp[-\frac{3}{2}H(t-t_0)]Z_{3/2}\left(\frac{\sqrt{-k}}{Ha_0}\exp[H(t-t_0)]\right)
+C_2\exp[-\frac{3}{2}H(t-t_0)]Z_{-3/2}\left(\frac{\sqrt{-k}}{Ha_0}\exp[H(t-t_0)]\right)-1
\end{equation}

\noindent where integral constant $C_1$ and $C_2$ can also be
fixed by initial values $\delta_0$ and $\dot{\delta_0}$. Using the
asymptotic expressions of Bessel functions and modified Bessel
functions at $x \gg 1$, we find that the anisotropy will increase
rapidly if the expansion rate $a(t)$ is slower than $a(t)\sim
t^{1/3}$ for $k=-1, 0$, or $a(t)\sim t$ for $k=+1$. Therefore,
during the matter dominated era and radiation dominated era, the
open and flat universe are stable against the spatially
homogeneous anisotropic perturbation during the matter and
radiation dominated era, while the closed universe is not and the
anisotropy increase exponentially.

 The recent data from WMAP indicate that the universe is almost flat\cite{ten,stein}.
In the following, we will rigorously prove the conclusion in the
flat universe case. The Eq.(\ref{deltaeq}) can be rewritten as

\begin{equation}\label{ineq}
\frac{\dot{\delta}(t)}{\dot{\delta_0}}=\left[\frac{a_0}{a(t)}\right]^3
\end{equation}

The scale factor $a(t)$ is determined by the pressure and energy
density of the perfect fluid in the Einstein equations
(\ref{einst1})-(\ref{einst3}). For example, we can apply these
equations specifically to the initial stage of cosmological
evolution which is assumed to be governed by the ordinary scalar
field. In this paper, these equations are applicable to matter
with energy momentum tensor of arbitrary form so that we should
discuss the varied form of $a(t)$. From Eq.(\ref{ineq}), we have
\begin{equation}\label{newineq}
 \delta_0+|\dot{\delta_0}|\cdot |\int_{t_0}^{t}\Big[\frac{a_0}
 {a(t)}\Big]^3dt|\leq\delta(t)\leq\delta_0+|\dot{\delta_0}|\cdot |\int_{t_0}^{t}\Big[\frac{a_0}
 {a(t)}\Big]^3dt|
\end{equation}

\noindent Next, we discuss the convergence of definite integral in
Eq.(\ref{newineq}). At $t\gg t_0$, we rewrite $a(t)$ as
$\frac{f(t)}{t^q}$, if $q>\frac{1}{3}$ and $f(t)\leq constant \leq
+\infty$, then the integral is convergent; If $q\leq\frac{1}{3}$
and $f(t)>constant\geq 0$, then the integral is divergent. This
argument can be extended to the following form by using the
consensus Cauchy criterion: if
$\bigg[\frac{a(t_0)}{a(t)}\bigg]^3=f(t)\cdot g(t)$ for $t\gg t_0$,
$\int_{t_0}^{\infty}g(t)dt$ is a convergent integral and $f(t)$ is
a monotonic and boundary function, then $I\equiv
\int\bigg[\frac{a(t_0)}{a(t)}\bigg]^3dt$ is a convergent integral.
Therefore, we have rigorously proved that flat universe is stable
against spatially homogeneous and anisotropic perturbation of
Kantowski-Sachs-type during the radiation and matter dominated
era.

Now, let summarize the key points of the above discussions. In
this paper, we focus on the Kantowski-Sach-like model of cosmology
and discuss the evolution of the cosmological anisotropy. We
showed that for the open and flat universe, the expansion rate of
the scale factor must be greater than $a(t)\sim t^{1/3}$ so that
the cosmological anisotropy does not increase while this
constraint becomes $a(t)\sim t$ for the closed universe. The
implication of these constraints are that: (i)Isotropic universe
must also be open or flat if it underwent the radiation and matter
dominated era in its past evolution. (ii)The oscillation universe
and static universe, corresponding to $q=0$, are generally
unstable against the spatially homogeneous anisotropic
perturbation unless the unnatural fine tuning is introduced. For
example, fine tuning $1+\delta_0+a_0\dot{\delta_0}=0$ is needed
for the static universe model. (iii) Together with the recent
observation, which favors closed and flat universe, it will be
better to say that the universe is flat.

As is known to all that when constructing spatially homogeneous
cosmological models, we simply choose a three-dimensional Lie
group \textit{G}, choose a basis of left invariant dual vector
fields on \textit{G} and choose a time-dependent left invariant
metric $h_{ab}(t)$ on \textit{G}. Then we can express the
spacetime metric $g_{\mu\nu}$ in term of $h_{ab}$\cite{wald}. The
Kantowski-Sachs-like model considered here can be treated by a
similar technique\cite{ryan}. Although there are many other
spatially homogeneous cosmological models that do not isotropize
and its cosmological anisotropy evolution may also be interesting
to study, they are beyond the scope of this paper.

 \vspace{0.8cm} \noindent ACKNOWLEDGMENTS

The authors thank P. Steinhardt and J. Barrow for helpful
comments. This work was partially supported by National Nature
Science Foundation of China under Grant No. 19875016 and
Foundation of Shanghai Development for Science and Technology
under Grant No. JC 14035.

\end{document}